# STM Study of Quantum Hall Isospin Ferromagnetic States of Zero Landau Level in Graphene Monolayer

Si-Yu Li[1], Yu Zhang[1], Long-Jing Yin[1,2], and Lin He[1,*]

[1] *Center for Advanced Quantum Studies, Department of Physics, Beijing Normal University, Beijing, 100875, China*

[2] *School of Physics and Electronics, Hunan University, Changsha, 410082, China*

[*]Correspondence and requests for materials should be addressed to L.H. (e-mail: helin@bnu.edu.cn).

**A number of quantum Hall isospin ferromagnetic (QHIFM) states have been predicted in the "relativistic" zero Landau level (LL) of graphene monolayer. These states, especially the states at LL filling factor $\nu = 0$ of charge-neutral graphene, have been extensively explored in experiment. To date, identification of these high-field broken-symmetry states has mostly relied on macroscopic transport techniques. Here, we study splitting of the zero LL of graphene at partial filling and demonstrate a direct approach by imaging the QHIFM states at atomic scale with a scanning tunneling microscope. At half filling of the zero LL ($\nu = 0$), the system is in a spin unpolarized state and we observe a linear magnetic-field-scaling of valley splitting. Simultaneously, the spin degeneracy in the two valleys is also lifted by the magnetic fields. When the Fermi level lies inside the spin-polarized states (at $\nu = 1$ or -1), the spin splitting is dramatically enhanced because of the strong many-body effects. At $\nu = 0$, we direct image the wavefunctions of the QHIFM states at atomic scale and observe an interaction-driven density wave featuring a Kekulé distortion, which is responsible for the large gap at charge neutrality point in high magnetic fields.**

In strong magnetic fields, the fourfold spin-valley invariance symmetry in graphene monolayer is frequently broken, giving rise to the SU(4) quantum Hall isospin ferromagnetism (QHIFM)[1-26]. The zero Landau level (LL) of graphene monolayer, which has no semiclassical analogue, is particularly interesting. The wave functions of each valley in the zero LL resides solely on one of the real-space sublattices, which plays a vital role in affecting the QHIFM states. Great efforts have been made to detect the high-field broken symmetry states of the zero LL and explore their exact nature[12-26]. Although enormous progress has been made, the QHIFM in graphene has so far been explored using transport techniques, which lack spatial resolution.

Theoretically, the wavefunctions of different phases in a QHIFM state, for example the predicted ground phases at half filling of the zero LL ($\nu = 0$), could be remarkably different[1-7]. This enables us to pinpoint the precise nature of the QHIFM state by directly imaging the wavefunctions in real space. Experimentally, the atomic structure of single graphene layer is fully accessible for direct imaging and the scanning probe techniques allow us to achieve atomic-scale spatial resolution and high-energy resolution. Thus, we study the QHIFM states of the zero LL in graphene monolayer and visualize their wavefunctions with a scanning tunneling microscope (STM). Our result indicates that both the valley and spin degeneracies in the zero LL are fully lifted in high magnetic fields. At $\nu = 0$ of charge-neutral graphene, a Kekulé distortion that triples the size of the unit cell of graphene, which is the characteristic feature of partially sublattice polarized (PSP) phase predicted at $\nu = 0$[7,9,16,25], is directly imaged in real space. Such a result indicates that the PSP phase is the precise nature of the QHIFM state at $\nu = 0$ of our graphene sample.

In our work, graphene multilayer samples were synthesized on Rhodium foils via a chemical vapor deposition (CVD) method[27,28]. The topmost graphene layer, which electronically decouples from the underlying graphene sheets through the large twist angle, behaves as a free-standing graphene monolayer[29-33]. We carried out STM measurements in defect-free regions to detect intrinsic electronic properties of pristine graphene and the area of a typical defect-free region can reach $1\times10^5$ nm$^2$ (see Fig. S1[34] for a typical large-area defect-free graphene monolayer). Figure 1(a) shows a representative graphene monolayer region decoupled from the underlying graphene layers with a twist angle of about 13.7 °(see Fig. S2 for the details[34]). Figure 1(b) shows two representative scanning tunneling spectroscopy (STS) spectra of the sample. In the presence of high magnetic field, well-defined LLs of massless Dirac Fermions are clearly observed (see Fig. S3 for Landau quantization of the topmost graphene layer measured in different magnetic fields[34]), as expected in a pristine graphene monolayer.

In the magnetic field of 9 T, the Fermi level crosses the zero LL of graphene and the zero LL is split (Fig. 1(b)), indicating that the valley and/or spin degeneracy in the zero LL are lifted. To clearly show this, we measured the spectra of the zero LL with high energy resolution of about 1 meV (see Supplementary materials for details[34]). Figure 1(c) shows a representative high-resolution tunneling spectrum of the zero LL recorded in 9 T, which reveals remarkable splittings. The zero LL splits into four peaks, indicating that both the spin and valley degeneracies are lifted[30], and the Fermi level exactly lies at $\nu = 0$: two peaks of the zero LL are filled while the other two peaks are empty. To further explore the nature of the splittings, we measured the high-resolution $dI/dV$ spectra of the zero LL as a function of magnetic fields (see Fig. S4[34]). For clarity, Fig. 1(d) shows differential tunneling conductance, i.e., $d^2I/dV^2$, of the zero LL measured in different magnetic fields. Each panel in Fig. 1(d) contains 8 $d^2I/dV^2$-$V$ spectra measured at different positions in a fixed magnetic field. Several observations can be made from the spectral prints. First, the zero LL splits starting from 2 to 3 T (see Fig. S4[34]), indicating the high sample mobility of the topmost graphene sheet in our experiment, which is vital to see the QHIFM states in graphene[2]. Second, the position of the zero LL shifts slightly with increasing the magnetic fields, indicating that the magnetic fields change the charge transfer between the topmost graphene layer and the supporting substrate. Similar phenomenon was also observed in previous STM studies in graphene monolayer supported by graphene multilayers[29,30,35,36]. In our experiment, the filling factor $\nu$ decreases from 2 (where the zero LL is fully filled) in the low magnetic fields to -1 (where only one of the four split peaks in the zero LL is filled) in the high magnetic field (see Fig. S4 for the high-resolution $dI/dV$ spectra at $\nu = 2, 1, -1$). The density of electron in the topmost graphene sheet versus the magnetic fields is easily calculated by $n = \nu B/\Phi_0$ ($\Phi_0$ is the flux quantum)[30], as shown in Fig. 1(e). Third, the splittings of the zero LL depends sensitively on both the magnetic fields and the filling factor, which enables us to explore the nature of the splittings, which will be elaborated subsequently.

We attribute the largest splitting of the zero LL to broken valley degeneracy (defined as $\Delta E_V$) and the other two splittings to broken spin degeneracy in the two valleys (defined as $\Delta E_{SL}$ and $\Delta E_{SR}$), as shown in Fig. 1(c). Figure 2 summarizes the obtained valley splitting and spin splitting as a function of magnetic fields and filling factors. In a previous work, tilted field magnetotransport measurements demonstrated explicitly that the insulating gap at $\nu = 0$ arises from broken valley degeneracy of the zero LL and the graphene system at $\nu = 0$ is spin unpolarized[16]. In our experiment, the valley splitting at $\nu = 0$ almost exhibits a linear scaling with magnetic field $B$ (Fig. 2(a)) and we obtain an effective $g$-factor of $g_v \approx 28$ with assuming a Zeeman-like dependence $E = g\mu_B B$ ($\mu_B$ is the Bohr magneton) or the slope for the splitting as 1.63 meV/T. Previous transport measurements[16] also observed a linear dependence of the valley splitting at $\nu = 0$ of the zero LL on $B$ and the measured effective $g$-factor is about $g_v = 23$. In the magnetic field of 9 T (the filling factor is 0), the valley splitting obtained in our experiment is about 16 meV, consistent with the insulating gap obtained at $\nu = 0$ of clean, high mobility graphene monolayer in previous transport measurements[12-16]. In graphene monolayer, the Coulomb energy is compatible with the measured valley splitting at $\nu = 0$. However, at present, the approximately linear $B$-scaling of the valley splitting in graphene is beyond the description of any existed theory[16,37] and more theoretical works are needed to fully understand this phenomenon. Moreover, our experiment demonstrated that the valley splitting depends sensitively on the filling factor: for example, the valley splitting decreases about 25% at $\nu = -1$ even though the magnetic field is higher (Fig. 2(a)). Such a behavior cannot be detected in transport measurements, which are only sensitive to energy scales of a few meV from the Fermi level (it is only possible to deduce the gap where the Fermi level crosses in transport measurements). The obvious decrease of the valley splitting from $\nu = 0$ to $\nu = -1$ indicates that exchange interaction plays an important role in the large valley splitting of the zero LL.

There are two other regions, as shown in Fig. 2(a), where the obtained valley splittings obviously deviate from the linear $B$-scaling. The first region is in the magnetic fields between 10 T and 11 T. Carefully examining the STS spectra measured between 10 T and 11 T reveals that the zero LL splits into five peaks (see Fig. S5[34]), suggesting the emergence of new states at fractional fillings. Similar splitting has been reported in a STM study of the $N = 1$ LL of graphene monolayer on top of graphene multilayers and the splitting at fractional fillings was attributed to the formation of counterflow superfluid states[30]. The emergence of new states at fractional fillings may result in the enhanced valley splitting between 10 T and 11 T observed in our experiment. To further explore the precise nature of the new state at fractional fillings, ultralow-temperature STM with ultrahigh energy resolution will be needed in the future. The second region is in the magnetic field below 5 T, where the zero LL is fully filled ($\nu = 2$) and the valley splitting depends weakly on the magnetic fields (Fig. 2(a)). Such a behavior indicates that the observed valley splitting below 5 T may mainly arise from sublattice symmetry breaking[16,25,38]. The supporting substrate can introduce on-site potential difference between the two sublattices of graphene, which breaks the inversion symmetry and lifts the valley degeneracy of graphene.

Besides the large valley splitting, the magnetic fields also lift the spin degeneracy in the two valleys of the zero LL, as shown in Fig. 2(b). At $\nu = 0$, the spin splittings for the fully filled valley and empty valley are almost the same and increase linearly with the magnetic fields (Fig. 2(b)). It is quite reasonable that the spin splittings of the two valleys exhibit the same linear $B$-scaling. Here, we should point out that the spin splittings at $\nu = 0$ of the zero LL cannot be detected in transport measurements because the Fermi level does not cross the gap between the spin-up and spin-down states. The effective $g$-factor, derived from the slope of the linear fit, is about 9, which is much larger than that for electron spins in graphene ($g = 2$). In previous transport measurements, it was demonstrated that the $g$-factor for spin splitting increases with decreasing LL index in graphene monolayer: the effective $g$-factor for spin splitting is measured to be 4 for the -3 LL, 6 for the -2 LL, and it increases to about 7 for the -1

LL[16]. Therefore, the measured effective *g*-factor ~ 9 in the zero LL is quite reasonable. Such a result indicates that strong electron-electron interactions in the presence of high magnetic fields may enhance the spin splitting in graphene monolayer beyond the normal Zeeman effect. The observed large effective *g*-factor for spin splitting is attributed to flip multiple spins smoothly in a skyrmionic spin texture[16]. Our experiment also demonstrates that the spin splitting depends sensitively on the filling factors. When the Fermi level lies inside the spin-polarized states (at ν = 1 or -1), we obtain an obvious enhanced spin splitting, which can exceed about 200% at maximum (Fig. 2(b)). Similar large enhancement of the spin splitting has also been observed in the *N* = 1 LL of graphene monolayer[30], which is attributed to the effects of exchange interaction.

After identifying the origin of the splittings in the zero LL of graphene, we will show subsequently that it is facile to directly image their local density of states (LDOS) at atomic scale with a STM. The LDOS at position *r* is determined by the wavefunctions according to $\rho(r) \propto |\psi(r)|^2$. Below we focus on possible quantum phases at the ν = 0 QHIFM state. Theoretically, four quantum phases, including fully spin-polarized ferromagnetic (F) state, spin-singlet charge-density-wave (CDW) state, spin-singlet PSP state characterized by Kekulé distortion, and canted antiferromagnetic (CAF) state, are predicted for the ν = 0 QHIFM state in graphene[2-8]. Previous transport measurements demonstrated that the ν = 0 QHIFM state in graphene is a spin un-polarized state[16], which help us to rule out the F state. However, direct distinguish the CDW, PSP, and CAF states through transport measurements is a very big challenge in experiment[16,21,25,26]. Figure 3(a) shows the atomic-scale wavefunctions of the three phases, which exhibit different features at atomic scale. At ν = 0, one spin-up and one spin-down states in a valley of the zero LL are filled. Therefore, the LDOS of the filled spin-up (or spin-down) state should exhibit triangular lattice for both the CDW and the CAF phases. The difference between the CDW phase and the CAF phase is that the LDOS of the spin-up and spin-down states locate at the same triangular lattice for the CDW phase, whereas they locate on different triangular lattice of graphene for the CAF phase. For the PSP phase, the LDOS of both the spin-up and spin-down states exhibit the characteristic Kekulé distortion, as represented schematically in Fig. 3(a).

By taking advantage of the atomic-scale spatial resolution of scanning probe techniques, we can direct image the ground state of the $\nu = 0$ QHIFM state, which helps us to pinpoint the precise nature of the QHIFM states in our sample. We imaged the QHIFM state by operating energy-fixed STS mapping, which reflects the LDOS in real space. Figure 3(b) and 3(c) show two representative STS maps recorded at different energies by fixing the filling factor $\nu = 0$. For the STS map recorded at the energy of -1 LL (Fig. 3(b)), we observe well-defined hexagonal honeycomb lattices. Such a result is quite reasonable because that the wavefunctions of each valley for the nonzero LLs are spreading equally on the two graphene sublattices. When measured STS maps at one of the split peaks of the zero LL, we observe a clear C-C bond-density wave in the carbon honeycomb network (Fig. 3(c)), which is the characteristic feature of the Kekulé distortion in the PSP phase (see Fig. S6 for more experimental data measured at the split peaks of the zero LL[34]). The Kekulé distortion triples the unit cell of graphene and breaks its chiral symmetry, which concomitantly opens an energy gap at the Dirac point of graphene[39-41]. Therefore, our experiment indicates that strong electron-electron interactions in high magnetic fields lead to the observed Kekulé distortion, which concomitantly generates the large gap at $\nu = 0$ of graphene. Interestingly, the C-C bond-density waves of the Kekulé distortion are so strong that we can direct image them even in the STM topographic measurement, as shown in Fig. 3(d). Theoretically, with considering the effects of magnetic fields on the Kekulé distortion may help to fully understand the approximately linear $B$-scaling of the gap, i.e., the valley splitting, at $\nu = 0$ of graphene.

In summary, we study the QHIFM states of the zero LL in graphene and visualize the wavefunctions of the QHIFM states with a STM. The magnetic fields lift both the valley and spin degeneracies in the zero LL, giving rise to the QHIFM states in graphene. At $\nu = 0$, interaction-driven density wave featuring a Kekulé distortion is directly imaged, indicating that the Kekulé distortion is the origin of the large gap at $\nu = 0$ of our graphene sample.

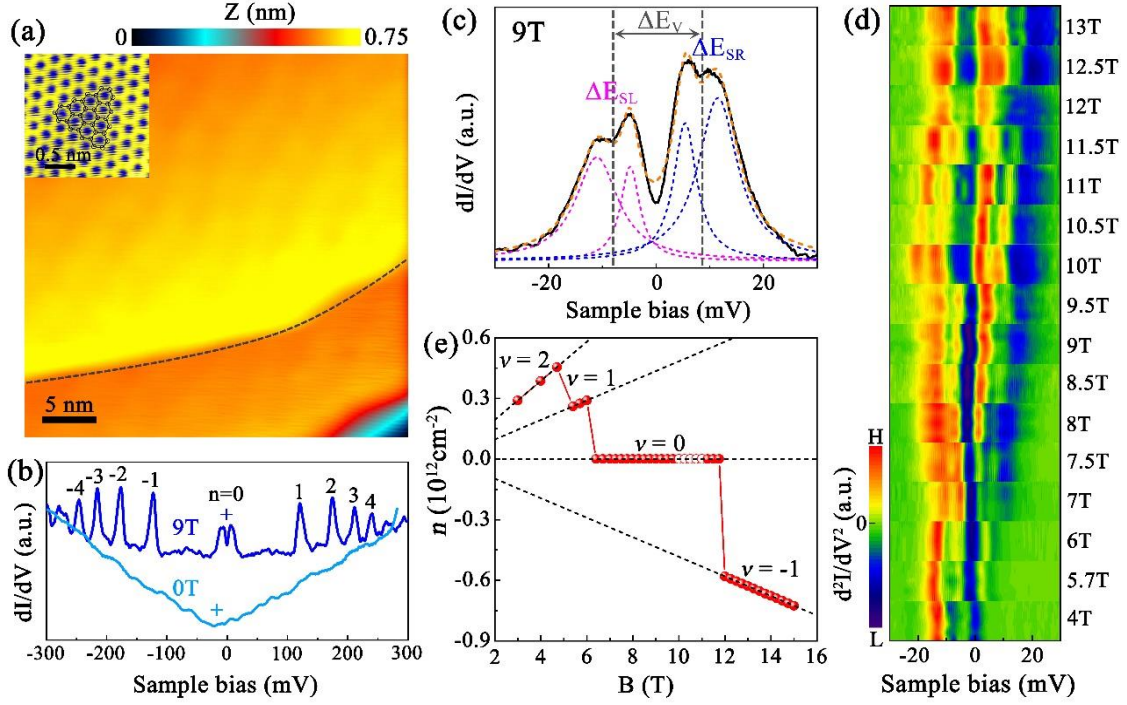

**FIG. 1.** (a) A 35 nm ×35 nm STM image ($V_{sample}$ = 1 V and $I$ = 0.4 nA) of graphene monolayer on Rh step. The color bar above shows the height fluctuation of this area. Inset: a 2 nm ×2 nm atomic resolved STM image ($V_{sample}$ = 0.9 V and $I$ = 0.4 nA) taken in the graphene monolayer with hexagonal honeycomb lattices. (b) The $dI/dV$-$V$ spectra acquired in panel (a) in 0 T and 9 T, respectively. The LL indexes and the positions of Dirac point in each spectrum are marked. (c) The high-resolution $dI/dV$-$V$ spectrum of the zero LL. The magenta and blue dashed peaks are the Lorentz fitting peaks for the splittings of the zero LL, and the orange dashed peak marks the superimposed result of the four fitting peaks. The energy separation of valley splitting $\Delta E_V$ and spin splittings $\Delta E_{SL}$, $\Delta E_{SR}$ are defined. (d) The $d^2I/dV^2$ spectra of the zero LL as a function of magnetic fields $B$. The color scale shows the relative intensity of the $d^2I/dV^2$. Each panel contains 8 $d^2I/dV^2$ spectra measured in each magnetic field. (e) The electron density of graphene monolayer $n = \nu B/\Phi_0$ versus the magnetic field $B$, where $\Phi_0$ is the flux quantum and $\nu$ is the filling factor.

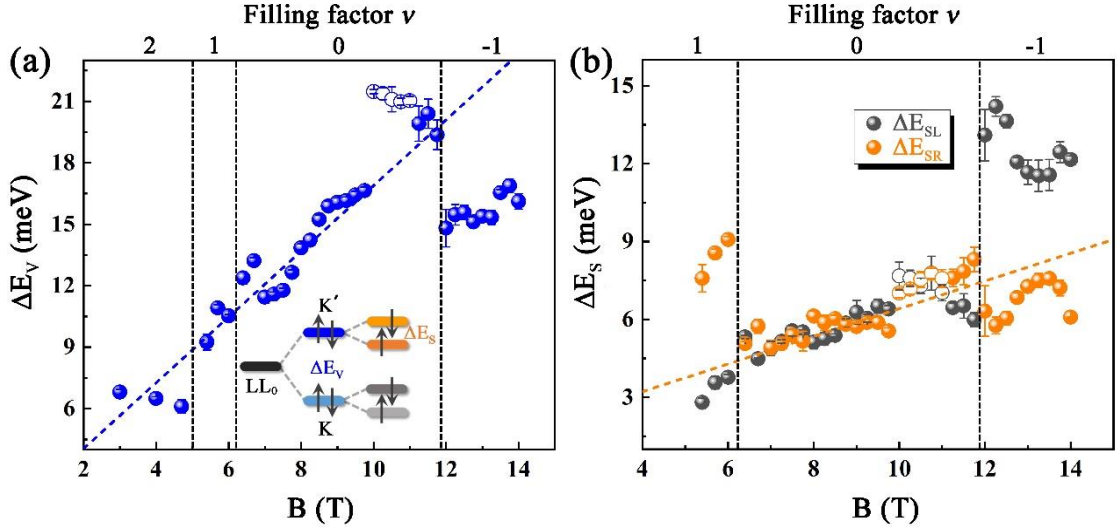

**FIG. 2.** (a) The valley splitting $\Delta E_V$ of the zero LL as a function of magnetic field. The blue dashed line marks the linear fitting for $\Delta E_V$ at $\nu = 0$, yielding the effective g-factor $g_V = 28$. The open circles between 10 T and 11 T correspond to the STS spectra that the zero LL splits into five peaks. Then the filling factor is hardly determined in our experiment. Inset, schematic of the energy level structure of the zero LL, showing the breaking of the valley symmetry followed by the breaking of the spin symmetry. (b) The spin splittings $\Delta E_{SL}$ and $\Delta E_{SR}$ versus the magnetic field. The linear fitting for the spin splitting at $\nu = 0$ yields the effective g-factor $g_S = 9$. When the Fermi level lies inside the spin-polarized states (at $\nu = 1$ or -1), the spin splitting is dramatically enhanced.

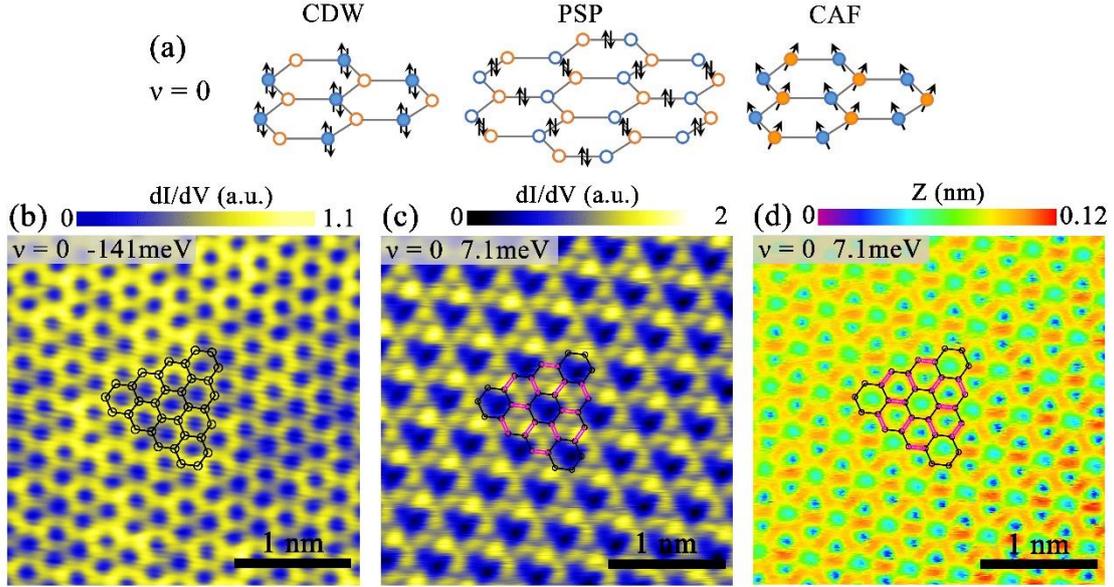

**FIG. 3. (a)** Schematics of the ground states with sublattice and spin polarization for the zero LL at $\nu = 0$, including the charge density wave (CDW) state, the partially sublattice polarized (PSP) state, and the canted antiferromagnetic (CAF) state. **(b)** A 3 nm ×3 nm STS map taken at -141meV corresponding to the energy of -1 LL. The filling factor of the sample is $\nu = 0$ and the magnetic field is 13 T. The honeycomb structures of graphene are overlaid onto the atomic resolved STS map. **(c)** A 3 nm ×3 nm STS map taken at the energy of 7.1 meV, which corresponds to the one of the empty peak in the zero LL. The filling factor of the sample is $\nu = 0$ and the magnetic field is 13 T. The honeycomb structures of graphene are overlaid onto the atomic resolved STS map. Clearly C-C bond-density wave, which is a characteristic feature of the Kekulé distortion in graphene, is observed. **(d)** A 3 nm ×3 nm STM image ($V_{sample}$ = 7.1 mV and $I$ = 0.08 nA) taken at $\nu = 0$. The Kekulé distortion is also clearly observed with comparing the inset schematic honeycomb lattices.